\documentclass[prd,twocolumn]{revtex4}
\usepackage{graphicx}
\usepackage{amssymb}
\usepackage{epstopdf}

\begin{document}

\title{Macroscopic Objects in Theories with Energy-dependent Speeds of Light}
\date{\today}

\author{Simon DeDeo}
\email{simon.dedeo@ipmu.jp}
\affiliation{Kavli Institute for Cosmological Physics, University of Chicago, Chicago, IL 60637, USA \\ \& \\ Institute for the Physics and Mathematics of the Universe, University of Tokyo, Kashiwano-ha 5-1-5, Kashiwa-shi, Chiba 277-8582, Japan}
\author{Chanda Prescod-Weinstein}
\email{cweinstein@perimeterinstitute.ca}
\affiliation{Department of Physics \& Astronomy, University of Waterloo, Waterloo, ON N2L 3G1, Canada \\ \& \\ Perimeter Institute, Waterloo, ON N2L 2Y5, Canada}

\begin{abstract}
Energy-dependent speeds of light have been considered an observable signature of quantum gravity effects. The two simplest dispersion relationships produce either linear or quadratic corrections, in particle energy, to the photon speed. The macroscopic limits of these theories -- how objects with small energy per particle, but with large mass, behave -- are not fully understood. We here briefly discuss some features of the macroscopic limit, that are necessary for understanding how detectors and emitters interact with the high-energy photons that probe spacetime.\end{abstract}

\maketitle

\section{Introduction}
\label{intro}

Theories that seek to unify the descriptions of General Relativity and Quantum Mechanics naturally incorporate the Planck energy-scale, $M_\mathrm{Pl}$, near which they make testable predictions. Energy (as opposed to the four-momentum vector) is a frame-dependent quantity and in a scattering experiment the most natural frame is that of the center-of-mass of the particles. Statements, then, that a particular quantum gravity effect might arise at energies close to the Planck scale are simplest to understand if the effect is the outcome of a local scattering process with a Planckian center-of-mass energy.

If we preserve Lorentz invariance in the standard fashion, however, very few astrophysical constraints after inflation will be competitive with either atomic physics or collider experiments. For example, the ``GZK cutoff'' processes~\cite{Greisen:1966p2247,Zatsepin:1966p2248}, due to photo-pion production on CMB photons by cosmic-rays, have a center-of-mass energy of approximately $100~\mathrm{MeV}$. Energies approaching the highest center-of-mass frames found in cosmic-ray collisions will soon be achieved, and probed at far greater precision, by the Large Hadron Collider~\cite{Ellis:2008p4163}.

This has not been the end of the story. A variety of suggestions have been made for how various astrophysical tests might provide new constraints on quantum gravitational effects. One, an energy-dependent speed of light, $c(E)$, has received great attention as a testable consequence of quantum gravity~\cite{AmelinoCamelia:1998p2181}. 

In this brief note, we examine the ``macroscopic limit'' of such theories, to determine how the constraints on macroscopic behavior might alter the theory on the one-particle level that has been the center of much attention. We discuss a toy model, which we call an ``observer preference'' theory, that can resolve the tension between $c(E)$ theories, standard macroscopic transformations, and laboratory results on the Doppler shift of light.

\section{$c(E)$ without preferred frames}
\label{cwpf}

It is possible to construct, for many situations, a consistent framework in which the speed of light may be energy-dependent and yet there be no detectable preferred-frame effects. A particularly important example is ``doubly special relativity'' (DSR)~\cite{AmelinoCamelia:2001p2304,AmelinoCamelia:2001p2416}.

Early on in studies of DSR~\cite{AmelinoCamelia:2002p9859} it was understood that the behavior of macroscopic objects -- \emph{i.e.}, objects composed of multiple particles -- could in many cases not be predicted by a simple extrapolation of the single-particle DSR laws. This includes both the theories termed (by Ref.~\cite{AmelinoCamelia:2002p9859}) ``DSR1''~\cite{AmelinoCamelia:2002p9781} and ``DSR2''~\cite{Magueijo:2002p11389}. It is possible have modified boost transformations that respect the Planck scale and have an ordinary macroscopic limit -- ``DSR3'' is given as one example in Ref.~\cite{AmelinoCamelia:2002p9859}; such theories require more elaboration to specify $c(E)$, which we do not undertake.

Here we approach the question from a different angle and investigate what happens if we allow $c(E)$ to vary, but maintain the standard, Lorentzian transformation laws of macroscopic objects. Such an approach is different from that of, for example, Ref.~\cite{AmelinoCamelia:2002p9781}, since rather than build up the two-particle (and higher) cases from the one-particle case, we impose a relationship in the large-$N$ limit. By describing how macroscopic objects perform as measuring devices, we can investigate the implications of unusual one-particle relationships between $c$ and $E$ for the multi-particle sector.

We begin by attempting to preserve the equivalence of inertial frames -- that the outcome of any local experiment should not depend on an observer's velocity. At the same time, we wish to preserve the metric nature of relativity and so accept the General Relativistic definition of inertia as established by, for example, the Pound-Rebka experiment: a particle that satisfies the unaccelerated geodesic equation is in such a frame.

Modifying transformation laws may make the notion of inertial frame ambiguous; one thread of recent work makes this issue explicit by introducing an energy-dependent metric~\cite{Smolin:2008p2498}. If different particles carry around different metrics, their associated geodesic equations may differ, and thus will their notion of inertial frame. Two particles moving relative to each other will disagree -- to order $E_{\mathrm{com}}/M_{\mathrm{Pl}}$ -- about who is truly accelerating.

Yet the phenomenologist need not worry about such issues, because the only observationally relevant inertial frames are those associated with macroscopic objects: stars, galaxies and satellites. Later formulations of DSR (\emph{e.g.}, that of Ref.~\cite{Magueijo:2003p2402}) suggested that these frames should -- roughly -- share the same notion of inertiality and furthermore be connected by the standard Lorentz transformations~\footnote{Such theories satisfy constraints on macroscopic-object Lorentz invariance by violating high-energy ``Gallelian'' intuitions: transformation laws (and thus, under a metric theory, gravitational couplings) may depend on the internal constituents of the object in question. A flea has a Planck-scale rest mass, but a transformation law very different from a Planck-energy electron, since the energy per particle of a flea is on the order of 1~GeV.}. 

According to Ref.~\cite{Magueijo:2003p2402}, the relevant quantity for determining the deformation of the transformation properties of an object is not total energy, but energy per elementary particle. Thus, the modification of the transformation laws for an ordinary object -- whose elementary particles appear as low-energy electrons and the constituents of protons and neutrons -- should be at most of order $m_\mathrm{proton}/M_\mathrm{Pl}$ -- or more than $10^{-3}$ less than for a TeV photon. 

This suggests, then, that assuming quantities associated with such ``macroscopic'' objects that serve as clocks and rulers transform in the usual Lorentz fashion is a reasonable starting point. Restricting our study to the interaction of macroscopic objects with individual high-energy particles is one way to avoid some of the paradoxes and difficulties of Ref.~\cite{Schutzhold:2003p473}.

Given these preliminaries, we can now formulate an effective theory for the measurement, by macroscopic objects, of energy dependent speeds of light. If $dx$ is the distance travelled by a photon (as measured by an inertial, macroscopic observer), and $dt$ is the time that observer measures, an energy dependent speed can be written formally as:
\begin{equation}
\label{ce}
\left(\frac{dx}{dt}\right)= c(n_{\mu} \bar{p}^\mu),
\end{equation}
where $(t,x,y,z)$ (a 4-vector, $x^\mu$) are the local co-ordinates of a macroscopic, inertial observer, $n^\mu$ is her 4-velocity, and $c(x)$ is some function. Since these two vectors (as well as the differential $dx^\mu$) are associated with measurements made by observers using macroscopic objects, we take them to transform in the standard fashion.

Conversely, we write $\bar{p}^\mu$ as the 4-momentum of the photon. This momentum is written with an overbar to indicate that it may transform differently from the vectors associated with macroscopic observers (and thus, \emph{e.g.}, that its index might be raised and lowered by a different metric.) Because of this, $n_\mu \bar{p}^\mu$ may not transform as a scalar and its value will be frame dependent.

For the sake of argument, we take $c(E)$ to be parametrized by slight departures at the Planck scale, \emph{i.e.},
\begin{equation}
c(E)=1-\alpha\left(\frac{E}{M_{\mathrm{Pl}}}\right)^n,
\end{equation}
where $\alpha$ is positive, and $n$ is positive (to have a hope of recovering the low-energy limit.) Our choice of units here amounts to setting the zero-energy speed of light, $c(0)$, and thus the parameter for macroscopic Lorentz transformations, equal to unity.

Given this choice, we can then ask if Eq.~\ref{ce} can be made consistent in all frames, given that $x^\mu$ (the coordinates of the macroscopic observer, frame $\mathcal{O}$) and $n^\mu$ (her 4-velocity) transform in the usual Lorentz fashion, but $\bar{p}^\mu$ may not.

Let us take a (macroscopic) primed frame, $\mathcal{O}^\prime$ to be moving in the negative $\hat{x}$ direction with velocity $v$ with respect to $\mathcal{O}$. In the classical case, we would expect the primed frame to observe a blueshift. Having no preferred frame allows that $E$ may go to $E^\prime$ in some strange fashion, as long as Eq.~\ref{ce} still holds. We then have

\begin{equation}
\label{newframe}
\frac{dx+v~dt}{dt+v~dx}=1-\alpha\left(\frac{E^\prime}{M_{\mathrm{Pl}}}\right)^n,
\end{equation}
where the left-hand side is found by Lorentz transforming $dx^\mu$, and the right-hand side comes from requiring Eq.~\ref{ce} to hold in the $\mathcal{O}^\prime$ frame, with $E$ allowed to transform to $E^\prime$ in a fashion we shall attempt to determine below.

For Eq.~\ref{newframe} to hold, then, we find the ``quantum gravity'' Doppler-shift law -- the relation between measured photon energy in boosted frames -- to be
\begin{equation}
\label{transform}
E^\prime = E\left[\frac{1-v}{1+v\left(1-\alpha\left[\frac{E}{M_{\mathrm{Pl}}}\right]^{n}\right)}\right]^{1/n}.
\end{equation}

The relationship requires that a photon experience a redshift when observed in the moving frame, contrary to the classical result of a blueshift when the observer is moving towards the source. In a previous draft of this paper (arXiv:0811.1999v1), it was claimed that the $n$ equal to two case is consistent with classical redshift relations (given our assumptions above regarding macroscopic transformations); this is incorrect.

\section{A Toy Observer-Preference Model}

An ``observer preference'' model is a toy model that attempts to reconcile the results of the previous section, which appear to require the unusual Doppler shift formula of Eq.~\ref{transform}, with observation, by allowing a subset of non-colocated observers to see the standard Doppler-shift relationship and relaxing the Eq.~\ref{ce} relationship to hold only for this observer class.

For example, a particle $A$ might emit a photon at point $x_A$ (and time $t_A$); at point $x_B$ (and time $t_B$), it might be absorbed by a second particle, $B$. We can enforce that particle $B$ measure the photon to have an energy $E_B$ related to $E_A$ by
\begin{equation}
\label{standard}
E_B=E_A\sqrt{\frac{1+v_0}{1-v_0}},
\end{equation}
where here we have assumed that particle $A$ has velocity $v_0$ in the direction of $B$ (\emph{i.e.}, for positive $v_0$, the photon is blueshifted in $B$'s frame.)

Doing so will mean that a global transformation, from the $A$ frame to the $B$ frame, that assumes standard macroscopic behavior will produce inconsistent results (the observed velocity-energy relationship of the photon will not be fixed by Eq.~\ref{ce} in all frames.) On the other hand, it will be consistent with laboratory measurements of Doppler shifts.

In order to see the effect on light propagation times, we will need to specify the energy of the photon along its entire path as measured by the ``observer class'' -- the collection of 4-vectors along the path that see the relationship of Eq.~\ref{ce} obtain. The only constraint on this class is that at the endpoints of the path, the observers match the velocity of the emitter (at point $A$) and absorber (at point $B$.)

As one choice of the arbitrary function that achieves this, we take
\begin{equation}
\label{interpolation}
v(\tau) = v_0\left(1-\frac{\tau}{\tau_F}\right),
\end{equation}
where $v(\tau)$ is the velocity of the observer class (in the reference frame of the absorber), $\tau$ is the (macroscopically measured) proper time along the photon path, $\tau_F$ is the proper time at the absorber, and $v_0$ is the velocity of the emitter.

At each point on the photon path,
\begin{equation}
\label{obsclass}
\frac{dx_\tau}{dt_\tau}=c(E_\tau),
\end{equation}
where the $\tau$ subscript indicates that we are in a frame comoving with observer at point $\tau$, and $E_\tau$ is the measured energy of the photon by this observer, which we take to be given by the standard Doppler shift relationship:
\begin{equation}
E_\tau = E_A\sqrt{\frac{1+v^\prime}{1-v^\prime}},
\end{equation}
where $v^\prime$ is the velocity of the emitter in the frame of the observer at $\tau$, given by the macroscopic relativistic formula
\begin{equation}
v^\prime = \frac{v_0-v(\tau)}{1-v_0v(\tau)}.
\end{equation}
Now requiring the velocity, fixed by the relationship Eq.~\ref{obsclass} in that frame only, to transform in the standard macroscopic fashion, we have
\begin{equation}
\frac{dx}{dt}=\frac{v(\tau)+c(E_\tau)}{1+v(\tau)c(E_\tau)},
\end{equation}
in the coordinate system of the absorber, which, to lowest non-zero order in $E$, and first order in $v_0$, is, for $n=1$,
\begin{equation}
\frac{dx}{dt}=1-\alpha\left(\frac{E}{M_\mathrm{pl}}\right)\left(1-v_0\left(2-3\frac{\tau}{\tau_F}\right)\right)+\cdots,
\end{equation}
and, for $n=2$,
\begin{equation}
\frac{dx}{dt}=1-\alpha\left(\frac{E}{M_\mathrm{pl}}\right)^2\left(1-v_0\left(2-4\frac{\tau}{\tau_F}\right)\right)+\cdots.
\end{equation}
These propagation speeds are different from those expected from the $c(E)$ relationship observed by $B$, who observes the photon blueshift in the standard fashion of Eq.~\ref{standard}. Only when the photon is at the absorber $B$ -- and thus $dx/dt$ coincides with the frame of the observer -- do the two quantities become equal. Different choices of observer class (Eq.~\ref{interpolation}) will produce different results; for example, a velocity observer class defined as a linear function of $\tau$, but in the frame of the emitter and not the absorber, produces a $dx/dt$ relationship that differs at $\mathcal{O}(v_0^3)$.

Using the fact that
\begin{equation}
\tau = \int \sqrt{1-\left(\frac{dx}{dt}\right)^2} dt,
\end{equation}
and that $\tau$ is a scalar quantity, we can solve for $\tau$ as a function of $t$ to find (to lowest order in $E$ and first order in $v_0$)
\begin{equation}
\label{n1}
\frac{dx}{dt}=1-\alpha\frac{E}{M_\mathrm{pl}}\left(1-2v_0+3\frac{t}{t_F}v_0+\ldots\right)
\end{equation}
in the $n=1$ case, and
\begin{equation}
\label{n2}
\frac{dx}{dt}=1-\alpha\left(\frac{E}{M_\mathrm{pl}}\right)^2\left(1-2v_0+4\frac{t}{t_F}v_0+\ldots\right)
\end{equation}
in the $n=2$ case.

This suggests that in some cases one could maintain standard macroscopic transformations and standard laboratory results, if one took an ``observer preference'' model. The cost, in this toy model, is that the $c(E)$ relationship holds only locally, and only for one particular member of an observer class defined by an emitter-absorber pair and an interpolation rule such as Eq.~\ref{interpolation}.

Note that the equation of motion for a photon, observed at $B$ to have energy $E_B$, but emitted by a stationary observer, is different from that of Eqs.~\ref{n1} and~\ref{n2}. In particular, since all members of the observer class are stationary relative to each other, we have (for the $n$ equal one case, \emph{e.g.}), and taking the photon energy to be $E\sqrt{(1+v_0)/(1-v_0)}$,
\begin{equation}
\frac{dx}{dt} = 1-\alpha\frac{E}{M_\mathrm{Pl}}(1+v_0+\cdots)
\end{equation}
which, in contrast to Eq.~\ref{n1}, has no apparent acceleration term -- showing explicitly that the photon path depends upon the relative velocity of the emitter and absorber (our derivations above are ``frame free'' in the sense that identical results are obtained if the absorber is taken to move towards the emitter.) Taken at face value, this leads to strangely non-local and potentially acausal effects: the path a photon takes depends upon the velocity of the observer that will, in the future, observe it.

One simple way to resolve these paradoxes of ``action at a distance'' is to simply declare the various observer classes ahead of time. Instead of the interpolation law of Eq.~\ref{interpolation} depending on the emitter and absorber alone, we could fill the universe with observers, making sure that $n^\mu$ chosen at every point on the path agrees with the frame of the observer at that point. 

Such an instantaneous, dynamical fixing of frame is related to ``dragged ether'' models such as that of Ref.~\cite{Afshordi:2008p671}; the forbidding of co-located observers with relative boosts is equivalent to the breakdown of the theory at stream-crossing~\footnote{The ``dragged ether'' we consider here is quite different from the ``partially dragged ether'' of Fresnel~\cite{MllerPedersen:2000p2837}; the latter was a ``pre-relativistic'' theory invented to explain the aberration of starlight~\cite{Ferraro:2005p2831} and can be tested (and ruled out) by the Michelson-Morley experiment alone.}. In general, before stream-crossing, tests of such models must sample different parts of the ``ether stream'' -- easy on astrophysical scales, but far harder terrestrially.

Depending on the scale of the dragging -- Michelson-Morely experiments do require the dragging to be operative on scales of order the Earth radius -- they can be constrained by experiments that look for the Sagnac effect~\cite{Anderson:1994p2871}. 

The relative velocities within the experimental apparatus must be large; the Michelson-Gale experiment~\cite{Mineur:1927p2833} was the first sensitive test. Modern-day experiments with such properties include ``round-the-world clocks'' of Hafele and Keating~\cite{Hafele:1972p3010} and studies using the M\"{o}ssbauer effect to check time-dilations in rotating frames~\cite{Turner:1964p3060}.

Achieving high accuracy with experiments that must incorporate large relative velocities is difficult; precision M\"{o}ssbauer studies sensitive to drag achieve internal motions of $10^{-6}c$ compared to the $10^{-3}c$ CMB-relative speeds sensitive to drift. In past work, Ref.~\cite{Sudarsky:2002p2107} noted that many astrophysical constraints on theories that amount to preferred frames are already beaten by laboratory studies, describing it as ``the ghost of Michelson-Morley coming back for revenge.'' While we leave a detailed analysis, of how different ether-dragging models may be restricted by terrestrial experiments, to later work, the ``ghost of Michelson-Gale'' may amount to lighter constraints.

\emph{Acknowledgments}. We thank Lee Smolin for helpful discussion. SD thanks the Perimeter Institute (Canada) and the Institute for the Physics and Mathematics of the Universe (Japan) for their hospitality while this work was undertaken. CPW thanks Sabine Hossenfelder, Jonathan Hackett, Joseph Henson and in particular Clifford Johnson for thought-provoking discussions. CPW is supported by Perimeter Institute for Theoretical Physics. Research at Perimeter Institute is supported by the Government of Canada through Industry Canada and by the Province of Ontario through the Ministry of Research and Innovation. 

\bibliographystyle{apsrev}

\end{document}